\documentclass[a4paper,11pt]{article}
\usepackage{jheppub}

\pdfoutput=1
\usepackage{hyperref}
\usepackage{graphicx}
\usepackage[utf8]{inputenc}
\usepackage{xspace}
\usepackage{amsmath,amssymb}
\usepackage{bbm}
\usepackage{xcolor}
\usepackage{booktabs}
\usepackage{subcaption}
\usepackage{mathtools}
\usepackage{empheq}
\usepackage{cancel}
\usepackage{multirow}

\DeclareFontFamily{U}{rcjhbltx}{}
\DeclareFontShape{U}{rcjhbltx}{m}{n}{<->rcjhbltx}{}
\newcommand{\cor}{\Delta{\cal S}^{\rm cor.}}
\newcommand{\uncor}{\Delta{\cal S}^{\rm uncor.}}
\newcommand{\rd}{{\rm d}}

\DeclareMathOperator*{\SumInt}{%
\mathchoice%
  {\ooalign{$\displaystyle\sum$\cr\hidewidth$\displaystyle\int$\hidewidth\cr}}
  {\ooalign{\raisebox{.14\height}{\scalebox{.7}{$\textstyle\sum$}}\cr\hidewidth$\textstyle\int$\hidewidth\cr}}
  {\ooalign{\raisebox{.2\height}{\scalebox{.6}{$\scriptstyle\sum$}}\cr$\scriptstyle\int$\cr}}
  {\ooalign{\raisebox{.2\height}{\scalebox{.6}{$\scriptstyle\sum$}}\cr$\scriptstyle\int$\cr}}
}

\newcommand\LTeq{\mathrel{\stackrel{\makebox[0pt]{\mbox{\normalfont\tiny L.T.}}}{=}}}

\title{The analytic two-loop soft function for leading-jet \boldmath $p_T$}

\author[a,b]{Samuel Abreu,}
\author[c]{Jonathan R.~Gaunt,}
\author[a]{Pier Francesco Monni,}
\author[d]{Robert Szafron}

\affiliation[a]{CERN, Theoretical Physics Department, CH-1211 Geneva
  23, Switzerland}
\affiliation[b]{Higgs Centre for Theoretical Physics, School of Physics and Astronomy,
 The University of Edinburgh, Edinburgh EH9 3FD, Scotland, United Kingdom}
\affiliation[c]{Department of
  Physics and Astronomy, University of Manchester, Manchester M13 9PL,
  United Kingdom}
\affiliation[d]{Department of Physics, Brookhaven National Laboratory,
Upton, N.Y., 11973, U.S.A.}

\emailAdd{samuel.abreu@cern.ch}
\emailAdd{jonathan.gaunt@manchester.ac.uk}
\emailAdd{pier.monni@cern.ch}
\emailAdd{rszafron@bnl.gov}

\preprint{CERN-TH-2022-059}
\abstract{
  We present the calculation of the two-loop soft function for the
  transverse momentum distribution of the leading jet produced in
  association with any colour-singlet system (e.g.~a Higgs or a $Z$
  boson). This constitutes a central ingredient for the resummation of
  the above distribution as well as the jet-vetoed cross section at
  the next-to-next-to-next-to-leading logarithmic order, both of which play
  an important role in the precision physics programme at the Large
  Hadron Collider.
  The calculation is performed in soft-collinear effective theory with
  an appropriate regularisation of the rapidity divergences that occur
  in the phase-space integrals.
  We obtain analytic results by employing an exponential regulator and
  by taking a Laurent expansion in the jet radius $R$.
  All expressions are presented as ancillary files with this article.}

\begin{document}
\setlength{\parskip}{0pt}
\maketitle
\flushbottom


\section{Introduction}
\label{sec:intro}

Precise tests of the Standard Model at the LHC involving the production of colourless final states require accurate control over QCD radiation. Jet vetoes are commonly used to reduce the unwanted QCD background by rejecting, or vetoing, events containing an energetic jet, i.e., with transverse momentum exceeding some cut-off value $p_T^{\rm veto}$. The presence of the cut-off, typically much smaller than the hard scale for the process $Q$, leads to large logarithmic terms of the form $\alpha_s^n \ln^k (p_T^{\rm veto}/Q)$, with $k \leq 2n$, in the perturbative series of the jet-vetoed cross section. Such corrections can spoil the convergence of the perturbative expansion and thus require all-order resummation. The first resummation of such logarithms was performed in Ref.~\cite{Banfi:2012yh} at the next-to-leading logarithmic (NLL) accuracy, and subsequently at next-to-next-to-leading logarithmic (NNLL) accuracy in Ref.~\cite{Banfi:2012jm} using a direct calculation in QCD, as well as in Refs.~\cite{Becher:2013xia,Stewart:2013faa} in the framework of soft-collinear effective field theory (SCET)~\cite{Bauer:2000yr,Bauer:2001yt,Bauer:2002nz,Beneke:2002ph,Beneke:2002ni} by means of renormalisation group methods. The above NNLL calculations also include a numerical extraction of the ${\cal O}(\alpha_s^2)$ constant terms relative to the Born (often referred to as NNLL$^\prime$), whose analytic calculation will be the focus of the study initiated by the present paper.

A great variety of phenomenological applications for various colour-singlet processes at the LHC has followed these results.
In the important area of Higgs physics, state of the art predictions for the jet-vetoed (zero-jet) cross section based on the combination of NNLL resummation with a next-to-next-to-next-to-leading-order (N$^3$LO) fixed-order calculation have been presented in Ref.~\cite{Banfi:2015pju}, including an account of heavy-quark-mass effects~\cite{Banfi:2013eda} and the resummation of logarithms of the jet radius of the type $\alpha_s^n \ln^n R^2 $ following Ref.~\cite{Dasgupta:2014yra}.
Similarly, predictions for the exclusive one-jet cross section have been presented in Ref.~\cite{Liu:2012sz}, and combined with the zero-jet bin in Ref.~\cite{Boughezal:2013oha}.
More recently, further technology has been developed for the calculation of differential distributions in the zero-jet bin~\cite{Monni:2019yyr}, as well as to incorporate the effect of rapidity cuts on jets~\cite{Michel:2018hui}.
An alternative definition of jet vetoes encoding a dependence on the rapidity of the vetoed jets has been presented and studied in Refs.~\cite{Gangal:2014qda,Gangal:2016kuo,Gangal:2020qik}.
Finally, applications to other colour-singlet production processes, such as the production of multi-boson electroweak final states, have been carried out in Refs.~\cite{Becher:2014aya,Moult:2014pja,Jaiswal:2014yba,Monni:2014zra,Dawson:2016ysj,Kallweit:2020gva}.

Going beyond NNLL accuracy is crucial for keeping up with the foreseen performance of the LHC experiments.
Previous considerations beyond NNLL were made in Ref.~\cite{Becher:2013xia}. A numerical calculation of the two-loop soft function was presented in Ref.~\cite{Bell:2020yzz}, albeit using a different regularisation scheme to what we consider here. A full N$^3$LL calculation remains, however, unavailable.
This paper initiates a systematic study of the higher-order terms for arbitrary colour-singlet processes.
The separation of scales achieved within SCET allows the formulation of a factorisation theorem that we take as a starting point.
The jet-veto cross section is obtained as the integral of the leading-jet transverse momentum distribution, which belongs to the class of SCET$_{\rm II}$ problems. These are affected by the so-called factorisation (or collinear) anomaly~\cite{BenekeFA,Becher:2010tm}, which is connected to the presence of rapidity divergences in the ingredients of the factorisation theorem. Such divergences are not regulated by the standard dimensional regularisation scheme and therefore an additional (rapidity) regulator must be introduced.
The introduction of this regulator leads to a more involved structure of the renormalisation group equations that govern the evolution of the factorisation's building blocks.
Various approaches have been proposed in the literature to restore factorisation and achieve a consistent resummation. In this article, we adopt the rapidity renormalisation group method~\cite{Chiu:2011qc,Chiu:2012ir}, which relies on the introduction of a new scale $\nu$, whose renormalisation flow allows one to consistently resum the logarithms associated with rapidity divergences.

This article presents the first analytic two-loop computation of one of the critical ingredients for N$^3$LL resummation of the logarithms $\ln (p_T^{\rm veto}/Q)$, namely the soft function.
The article is organised as follows:
In section \ref{sec:factorisation}, we review the factorisation theorem for the production of a colour-singlet with a veto on the transverse momentum of the leading jet and discuss the definition of the soft function in the presence of a rapidity regulator. Section \ref{sec:computation} discusses our computation of the two-loop soft function. The results and various checks are presented in section \ref{sec:results}, and the conclusions are given in section \ref{sec:conclusions}.


\section{Factorisation of leading-jet transverse momentum in SCET}
\label{sec:factorisation}\sloppy
In this section we describe the formalism for the jet-vetoed cross
section in colour-singlet production, which is defined as the integral
of the leading-jet transverse momentum distribution up to a cut
$p_{T}^{\rm veto}$.
As mentioned in the introduction, we adopt the formalism of the
rapidity renormalisation group~\cite{Chiu:2011qc,Chiu:2012ir} to deal
with the rapidity divergences, as opposed to the original treatment of
Ref.~\cite{Becher:2010tm} in which the exponentiation of the collinear
anomaly follows from the consistency of the EFT formulation.

Our starting point is the factorisation theorem for this observable in
SCET~\cite{Becher:2012qa,Becher:2013xia,Stewart:2013faa}. We consider the production of a generic colour-singlet system of squared invariant mass~$Q^2$, with a veto on
the transverse momentum of the leading jet, 
$p_{T}^{\rm jet} < p_{T}^{\rm veto}$. In the limit
$p_{T}^{\rm veto}\ll Q$, the cross section differential in the kinematics of the colour-singlet system is given by
\begin{align}
\label{eq:factorisation}
\frac{d\sigma(p_{T}^{\rm veto})}{d\Phi_{\rm Born}} &\equiv
                                                     \sum_{F=q,g}
                                                     |{ A}^F_{\rm
                                                     Born}|^2\,{\cal
                                                     H}^F(Q;\mu)\notag\\
  &\times \,{\cal B}^F_n(x_1,Q,p_{T}^{\rm veto},R^2;\mu,\nu) \,{\cal B}^F_{\bar n}(x_2,Q,
  p_{T}^{\rm veto},R^2;\mu,\nu)\,{\cal S}^F(p_{T}^{\rm veto},R^2;\mu,\nu)\,,
\end{align}
where $d\Phi_{\rm Born}$ denotes the full Born phase-space measure
including the flux factor, ${ A}^F_{\rm Born}$ is the Born
amplitude for the production of the colour-singlet system, and $\mu$
and $\nu$ denote the renormalisation and rapidity regularisation
scales, respectively. 
All quantities on the second line of Eq.~\eqref{eq:factorisation}
depend explicitly on the jet radius $R$.
The index $F$ indicates the flavour configuration of the initial
state, i.e. either $q\bar{q}$ ($F=q$) or $gg$ ($F=g$), and for
simplicity we will drop it when referring to the ingredients of the
factorisation theorem~\eqref{eq:factorisation}.
In Eq.~\eqref{eq:factorisation}, the hard function ${\cal H}$
describes the dynamics at large virtuality scales $\mu\sim Q$, and
therefore contains purely virtual contributions. It is defined as
the square of the matching coefficient between QCD and SCET, i.e.,
\begin{equation}
{\cal H}(Q;\mu) = |{\cal C}(Q;\mu)|^2\,.
\end{equation}
The two beam functions ${\cal B}_n$ and ${\cal B}_{\bar n}$ describe
collinear dynamics of radiation along the light-cone directions
$n^\mu$ and ${\bar n}^{\mu}$, which correspond to the beam directions. 
These are defined by matrix elements of
collinear operators in SCET.
Finally, the soft function ${\cal S}$ describes the dynamics of soft
radiation emitted off the two initial-state partons.

The resummation of the logarithms $\ln Q/p_{T}^{\rm veto}$ is achieved
by solving evolution equations for each of the functions in the
factorisation theorem~\eqref{eq:factorisation} between their canonical
scales and the two scales $\mu$ and $\nu$.
Specifically, the hard function satisfies the Renormalisation Group
Equation (RGE) related to that of the matching coefficient
${\cal C}(Q;\mu)$ (see e.g. Refs.~\cite{Becher:2009qa,Gardi:2009qi}
and references therein)
\begin{equation}
\frac{d}{d\ln\mu}  \ln {\cal C}(Q;\mu)= \Gamma_{\rm cusp}^F(\alpha_s(\mu))
\ln\frac{-Q^2}{\mu^2} + \gamma_{H}^F(\alpha_s(\mu))\,,
\end{equation}
where $\Gamma_{\rm cusp}^F$ and $\gamma_{H}^F$ are the cusp and hard
anomalous dimensions of the quark ($F=q$) or gluon ($F=g$) form factor in
the $\overline{\rm MS}$ scheme.
The boundary condition of the evolution is set at the canonical scale
$\mu=Q$.
Unlike the hard function, the beam functions ${\cal B}_i$ also depend
on the rapidity regularisation scale $\nu$, and satisfy a system of
evolution equations (see
e.g. Refs.~\cite{Becher:2013xia,Stewart:2013faa}). The first is the
RGE
\begin{equation}
\frac{d}{d\ln\mu}\ln {\cal B}_i(x,Q,p_{T}^{\rm veto},R^2;\mu,\nu) = 2\,\Gamma_{\rm cusp}^F(\alpha_s(\mu))
\ln\frac{\nu}{Q} + \gamma_{B}^F(\alpha_s(\mu))\,,
\end{equation}
where $\gamma_{B}^F$ is the collinear anomalous dimension, and the second
is the rapidity evolution equation
\begin{equation}
\frac{d}{d\ln\nu}\ln {\cal B}_i(x,Q,p_{T}^{\rm veto},R^2;\mu,\nu) = 2
\int_{p_{T}^{\rm veto}}^{\mu}\frac{d\mu^\prime}{\mu^\prime}\Gamma_{\rm
  cusp}^F(\alpha_s(\mu^\prime)) -\frac{1}{2} \gamma_\nu^F(p_{T}^{\rm veto},R^2)\,,
\end{equation}
where $\gamma_\nu^F$ denotes the observable-dependent rapidity
anomalous dimension.
The boundary condition for the joint $\{\mu,\nu\}$ evolution is set at
the canonical scales $\mu=p_{T}^{\rm veto}$ and $\nu=Q$.
Finally, the soft function ${\cal S}$ satisfies the system of evolution equations
\begin{align}
 \label{eq:soft_RGES} 
\frac{d}{d\ln\mu}\ln {\cal S}(p_{T}^{\rm veto},R^2;\mu,\nu) =& 4\,\Gamma_{\rm cusp}^F(\alpha_s(\mu))
  \ln\frac{\mu}{\nu} + \gamma_{S}^F(\alpha_s(\mu))\,,\notag\\
\frac{d}{d\ln\nu}\ln {\cal S}(p_{T}^{\rm veto},R^2;\mu,\nu) =&
-4\int_{p_{T}^{\rm veto}}^{\mu}\frac{d\mu^\prime}{\mu^\prime}\Gamma_{\rm
  cusp}^F(\alpha_s(\mu^\prime)) + \gamma_\nu^F(p_{T}^{\rm veto},R^2)\,,  
\end{align}
with canonical scales $\mu=\nu=p_{T}^{\rm veto}$.
In the framework of the rapidity renormalisation group, the dependence
on the rapidity anomalous dimension cancels between the soft and beam
functions, while the soft ($\gamma_{S}^F$) and collinear
($\gamma_{B}^F$) anomalous dimensions are related to the hard
anomalous dimension $\gamma_{H}^F$ by the RG invariance of the
physical cross section, that is
\begin{equation}
2\gamma_{H}^F+\gamma_{S}^F+2 \gamma_{B}^F=0\,.
\end{equation}

The resummation of the jet-vetoed cross section at N$^k$LL will
require the cusp anomalous dimension $\Gamma_{\rm cusp}^F$ up to $k+1$
loops, and the quantities $\gamma_{H}^F$, $\gamma_{S}^F$,
$\gamma_{B}^F$, $\gamma_{\nu}^F$ up to $k$ loops.
Finally, the boundary conditions of the RGEs introduced in this
section are needed up to $k-1$ loops. In this article, we focus on the
calculation of such boundary conditions for the soft function up to
the two-loop order, which, together with the two-loop beam functions
and the three-loop rapidity anomalous dimension $\gamma_{\nu}^F$, are
the missing ingredients to formulate the N$^3$LL computation.

Before we carry on with the computation of the soft function, a remark
is in order. The validity of the factorisation theorem in
Eq.~\eqref{eq:factorisation} at arbitrary logarithmic order has been
the subject of debate in the SCET literature. Specifically, while its
validity up to NLL is uncontroversial, going to higher logarithmic
orders requires some extra care. The authors of
Refs.~\cite{Tackmann:2012bt,Stewart:2013faa} argue that, starting from
NNLL, the existence of \textit{soft-collinear mixing} terms may break
the factorisation of the jet-veto cross section into soft and beam
functions, while the authors of Ref.~\cite{Becher:2013xia} disagree
with this statement and argue that the cancellation of such terms
requires the multipole expansion of the phase-space constraint in the
computation.
In this article, we do not discuss this matter any further as it is
not directly relevant for the results presented here. We will present
a discussion of such soft-collinear mixing terms in a future
publication.

\section{The soft function}
\label{sec:computation}
In this section we focus on the soft function, whose operatorial
definition reads:
\begin{equation}
\label{eq:soft_def}
{\cal S}(p_{T}^{\rm veto},R^2;\mu,\nu) = \frac{1}{d_F}{\rm Tr}\left\{\langle
  0 | Y_{n}^{\dagger} Y_{\bar n} \,{\mathbb M}(p_{T}^{\rm veto},R^2)\,
  Y_{\bar n}^{\dagger} Y_{n}| 0 \rangle\right \}\,,
\end{equation}
where the soft Wilson line $Y_n$ is defined in terms of the soft gauge
field $A^{(s)}_a$ as~\cite{Bauer:2001yt}
\begin{equation}
Y_{n}(x) = {\cal P}\exp\left\{i g_s\int_{-\infty}^0 ds\, n\cdot
    A^{(s)}_a(x+n s){\mathbb T}^a\right\}\,, 
\end{equation}
and analogously for $Y_{\bar n}$.
The colour charge operator ${\mathbb T}^a$ is understood to be in the
adjoint (fundamental) representation of $SU(3)$ for gluon (quark)
initiated processes. The trace in Eq.~\eqref{eq:soft_def} is performed
over colour indices and the normalisation factor $d_F$ denotes the
dimension of the representation, i.e.~$d_F=N_C^2-1$ for
gluon-initiated reactions and $d_F=N_C$ for quark-initiated ones.
The operator ${\mathbb M}(p_{T}^{\rm veto},R^2)$ acts on a given state
of $X_s$ soft particles $|X_s\rangle$ by applying a veto
$p_{T}^{\rm veto}$ on the final-state jets of jet radius $R$, obtained
with a generalised $k_T$ family of clustering algorithms.

To proceed we introduce a complete set of soft states in
Eq.~\eqref{eq:soft_def} and recast it as
\begin{equation}
\label{eq:soft_def_states}
{\cal S}(p_{T}^{\rm veto},R^2;\mu,\nu) = \frac{1}{d_F}\SumInt_{X_s} {\rm Tr}\left\{{\cal M}(p_{T}^{\rm veto},R^2)\,\langle
  0 | Y_{n}^{\dagger} Y_{\bar n} |X_s\rangle 
\langle X_s| Y_{\bar n}^{\dagger} Y_{n}| 0 \rangle\right \}\,,
\end{equation}
where ${\cal M}(p_{T}^{\rm veto},R^2)$ is the phase-space constraint, which
schematically reads
\begin{equation}
  \label{eq:soft_def_measure}
{\cal M}(p_{T}^{\rm veto},R^2) = \Theta(p_{T}^{\rm veto}-\max\{p_T^{\rm jet_i}\})
\Theta_{\rm cluster}(R^2)\,.
\end{equation}
$\Theta_{\rm cluster}(R^2)$ is the generic clustering condition
on the $X_s$ soft particles in the final state. This is defined
algorithmically for the generalised-$k_T$ family of jet algorithms
with distance measures
\begin{equation}
d_{ij} = \min\{k_{i\perp}^{2 p},k_{j\perp}^{2p}\} \left[(\Delta\eta_{i j})^2 +
(\Delta\phi_{i j})^2\right]\,,\quad d_{iB}=k_{i\perp}^{2 p}\,,
\end{equation}
with $p=-1$ for the anti-$k_T$~\cite{Cacciari:2008gp}, $p=0$ for the
Cambridge-Aachen~\cite{Dokshitzer:1997in,Wobisch:1998wt}, and $p=1$
for the $k_T$~\cite{Catani:1993hr} algorithm.
Here $k_{i\perp}$ denotes the transverse momentum of particle $i$ with
respect to the beam direction, and $\Delta\eta_{i j}$ and $\Delta\phi_{i
  j}$ are the relative rapidity and azimuthal angle between particles
$i$ and $j$, respectively.
The clustering condition $\Theta_{\rm cluster}(R^2)$ specifies how
particles are sequentially clustered according to the above distance
measures.
Its expression will be given below when discussing the calculation of
the soft function.

The computation can be decomposed into the sum of the soft function
for a reference observable and a correction factor. 
A similar approach for the calculation of
two-loop soft functions has been exploited also in
Refs.~\cite{Gangal:2016kuo,Bauer:2020npd}. 
Here, for the reference observable we take the
transverse momentum of the colour singlet system, which we denote by
$ {\cal S}_{\perp}(p_{T}^{\rm veto},\mu,\nu)$ (defined in
Ref.~\cite{Li:2016axz} for the regularisation scheme adopted here).
The correction factor
$\Delta{\cal S}(p_{T}^{\rm veto},R^2;\mu,\nu)$ accounts for the
effect of the clustering algorithm. The soft function
${\cal S}_{\perp}(p_{T}^{\rm veto},\mu,\nu)$ is well understood and
known up to $\mathcal{O}(\alpha_s^3)$~\cite{Li:2016axz,Li:2016ctv}.
We recompute it up to ${\cal O}(\alpha_s^2)$ in this article to
validate our approach. The jet-veto soft function is then:
\begin{align}\label{eq:DeltaS}
{\cal S}(p_{T}^{\rm veto},R^2;\mu,\nu)  = {\cal S}_{\perp}(p_{T}^{\rm veto},\mu,\nu) +\Delta{\cal S}(p_{T}^{\rm veto},R^2;\mu,\nu) \,.
\end{align}
This equation defines the remainder term
$\Delta{\cal S}(p_{T}^{\rm veto},R^2;\mu,\nu) $, whose evaluation at
the two-loop level is the main result of this article.

The remainder function $\Delta{\cal S}$ depends on the jet algorithm
and only contributes for two or more real emissions. This means that
the difference $\Delta{\cal S}$ starts at $\mathcal{O}(\alpha_s^2)$
and at this accuracy is purely determined by double-real diagrams.
The decomposition in Eq.~\eqref{eq:DeltaS}, together with the
definition~\eqref{eq:soft_def_measure} allows us to determine the
measurement function for the remainder term $\Delta{\cal S}$ as
\begin{align}\label{eq:DeltaJ}
\Delta {\cal M}(p_{T}^{\rm veto},R^2) \equiv \Theta(p_{T}^{\rm veto}-\max\{p_T^{\rm jet_i}\})
\Theta_{\rm cluster}(R^2)- \Theta\left(p_{T}^{\rm veto}-\left|\sum_{X_s}p_T^{\rm jet_i}\right|\right)\,.
\end{align}

\subsection{Regularisation of rapidity divergences}
We now discuss the regularisation procedure used to deal with the
rapidity divergences that feature in SCET$_{\rm II}$ problems.
We adopt the exponential regulator initially proposed in
Ref.~\cite{Li:2016axz}, which is defined by modifying the phase-space
integration measure for each real emission, such that
\begin{equation}
  \label{eq:regulator_def}
\prod_{i} d^d k_i \delta(k^2_i)\theta(k^0_i)\to  \prod_{i} d^d k_i \delta(k^2_i)\theta(k^0_i)\exp\left[ \frac{-e^{-\gamma_E}}{\nu}(n \cdot k_i + \bar{n} \cdot k_i)\right]\,.
\end{equation}
where $\nu$ denotes the rapidity regularisation scale discussed in
section~\ref{sec:factorisation}.  On the other hand, the integration
measure for virtual corrections remains unchanged. The role of the
regulator is to suppress the integrand as the light-cone components
become large, hence regulating the phase-space region affected by
rapidity divergences. The regularised soft function is obtained by
taking the Laurent expansion around $\nu \to +\infty$ after the
phase-space integrals over $k_i $ have been evaluated. 
In the expansion, one neglects terms of $\mathcal{O}(\nu^{-2})$.
The exponential regulator introduced above has several useful
features. In particular, it preserves non-abelian exponentiation, and
allows for the resummation of rapidity logarithms by means of
evolution equations given in section~\ref{sec:factorisation}, whose
solution is independent of the path chosen in the $(\mu,\nu)$ plane.
In the following sections we present the results for the one-loop and
two-loop soft functions, and discuss how the above regulator is dealt
with in the explicit computation.

\subsection{The renormalised one-loop soft function}
\label{sec:oneloop}
We define the perturbative expansion of the soft function as
\begin{align}
{\cal S} = \sum_{k=0}^{\infty}\left(\frac{\alpha_s}{4\pi} \right)^k  {\cal S}^{(k)} \,. 
\end{align}
The leading order result is
${\cal S}^{(0)}(p_{T}^{\rm veto},R^2;\mu,\nu) = {\cal
  S}^{(0)}_{\perp}(p_{T}^{\rm veto},\mu,\nu) =\Theta(p^{\rm veto}_T)=1
$. At $\mathcal{O}(\alpha_s)$, only a single emission of a gluon is
possible, in which case Eq.~(\ref{eq:DeltaJ}) simplifies considerably
\begin{align}
\Delta {\cal M}(p_{T}^{\rm veto},R^2)  = 0,\,
\end{align}
implying that 
\begin{align}
{\cal S}^{(1)}(p_{T}^{\rm veto},R^2;\mu,\nu)  = {\cal S}^{(1)}_{\perp}(p_{T}^{\rm veto},\mu,\nu)\,.
\end{align}
In fact, the above considerations trivially hold true for all
single-emission contributions to all orders in $\alpha_s$. By explicit
computation, and performing the renormalisation in the $\overline{\textrm{MS}}$
scheme, one finds the result
\begin{align}
 {\cal S}^{(1)}_{\perp}(p_{T},\mu,\nu) = - C_F \left[ 8
   \ln \left(\frac{p_T}{\mu }\right) \left(2 \ln \left(\frac{\mu }{\nu
   }\right)+\ln \left(\frac{p_T}{\mu }\right)\right)+\frac{\pi ^2}{3} \right]\,.
\end{align}

\subsection{The renormalised two-loop soft function}
\label{sec:twoloop}
At two loops, we start by computing
$ {\cal S}^{(2)}_{\perp}(p_{T},\mu,\nu)$ using an analytic
implementation of sector decomposition and the \texttt{Mathematica}
packages \texttt{HypExp}~\cite{Huber:2005yg} and
\texttt{PolylogTools}~\cite{Duhr:2019tlz}. Our results agree with the findings of
Refs.~\cite{Li:2011zp,Li:2016ctv}, which provides a first consistency
check on our computational approach.
For completeness, we include them in the ancillary
file \texttt{SoftFunctionPerp.m}. 
At this order we also get the first non-zero contribution to
$\Delta{\cal S}(p_{T},R^2;\mu,\nu)$, coming from double-real diagrams
describing the emission of either two soft gluons or a soft
quark-antiquark pair. The corresponding double-real squared amplitudes
were obtained in
Refs.~\cite{Campbell:1997hg,Dokshitzer:1997iz,Catani:1999ss} (see also
Refs.~\cite{Kelley:2011ng,Monni:2011gb,Hornig:2011iu} for calculations
in the context of SCET).
At two-loops, the $\mu$ dependence in $\Delta{\cal S}(p_{T},R^2;\mu,\nu)$
is entirely encoded in the strong coupling constant $\alpha_s(\mu)$,
while the $\nu$ dependence is induced by the exponential regulator of
which we will eventually consider the expansion around
$\nu\to +\infty$.

It is convenient to split the calculation of 
$\Delta{\cal S}^{(2)}(p_{T},R^2;\mu,\nu)$ into the sum of a \textit{correlated}
and an \textit{uncorrelated} term (see 
also e.g.~Refs.~\cite{Gangal:2016kuo,Banfi:2014sua}). We write
\begin{align}
\label{eq:split}
\Delta{\cal S}^{(2)} (p_{T},R^2;\mu,\nu) \equiv \cor(p_{T},R^2;\mu,\nu) + \uncor(p_{T},R^2;\mu,\nu)\,,
\end{align}
which follows from a similar decomposition of the squared amplitudes,
\begin{equation}
| A(k_1,k_2)|^2\equiv {\cal A}^{\rm cor.}(k_1,k_2)+{\cal A}^{\rm uncor.}(k_1,k_2)\,.
\end{equation}
The uncorrelated term ${\cal A}^{\rm uncor.}(k_1,k_2)$ encodes the
limit of the double-soft squared amplitude $|A(k_1,k_2)|^2$ for
relative rapidity
$\eta\equiv \Delta\eta_{12}=\eta_1-\eta_2\to \infty$.  This term is
proportional to the colour factor $C_R^2$, where $C_R=C_F$ or
$C_R=C_A$ for quark or gluon initiated processes.
Conversely, the correlated term encodes the remaining
contributions to $|A(k_1,k_2)|^2$, which vanish in the limit where the
two emissions are widely separated in rapidity. This term
can be further separated into two colour factors, $C_R C_A$
and $C_R T_R n_F$:
\begin{equation}\label{eq:corrSplit}
\cor(p_{T},R^2;\mu,\nu)=
\Delta{\cal S}^{(2)}_{C_R \,n_F \,T_R} (p_{T},R^2;\mu, \nu)+
\Delta{\cal S}^{(2)}_{C_R \,C_A} (p_{T},R^2;\mu, \nu)\,.
\end{equation}
The utility of the decomposition in Eq.~\eqref{eq:split}
lies in the fact that the two terms on the right-hand side require different 
treatments of the rapidity regulator.

We have performed two independent computations for 
$\Delta{\cal S}^{(2)}(p_{T},R^2;\mu,\nu)$, one analytical and one numerical, which we
discuss in more detail below. Before doing so, however, we note that 
$\Delta{\cal S}^{(2)}(p_{T},R^2;\mu,\nu)$
is finite in all infrared and collinear limits of the two-particle
phase space (because the r.h.s.~of Eq.~\eqref{eq:DeltaJ} vanishes in
these limits), and therefore the calculation can be directly performed
in $d=4$ space-time dimensions. In particular,
this implies that $\Delta{\cal S}^{(2)}(p_{T},R^2;\mu,\nu)$ does
not depend on $\mu$.

\subsubsection{Analytic computation of the two-loop soft function}
\label{sec:twoloop-ana}
We start by discussing the analytic calculation of
$\Delta{\cal S}^{(2)}(p_{T},R^2;\mu,\nu)$.
The momenta of the two real particles (either gluons or quarks) are
denoted by $k_i$, and we adopt the following parametrisation for the
phase space in the r.h.s. of Eq.~\eqref{eq:regulator_def}:
\begin{align}
k_i = k_{i\perp} \left(\cosh\eta_i\,,\cos\phi_i\,, \sin\phi_i, \sinh \eta_i \,\right),\quad i=1,2\,,
\end{align} 
in terms of the (pseudo-)rapidities $\eta_i$, the azimuthal angles
$\phi_i$, and the transverse momenta
$k_{i\perp}\equiv |\vec{k}_{i\perp}|$.
We then perform a change of variable in the parametrisation
of $k_2$,
\begin{align}\label{eq:coordinates}
\left\{k_{2\perp},\eta_2,\phi_2\right\}\rightarrow
  \left\{\zeta\equiv
  k_{2\perp}/k_{1\perp},\eta\equiv\eta_1-\eta_2,\phi
  \equiv \phi_1 - \phi_2\right\}\,,
\end{align}
in order to express its kinematics relatively to the ones of $k_1$. With this
change of variable, the measurement function~\eqref{eq:DeltaJ} takes
the simple form
\begin{align}\label{eq:DeltaJ-2L}
  \begin{split}
\Delta {\cal M}(p_{T}^{\rm veto},R^2) \equiv &\left[\Theta(p_{T}^{\rm veto}-k_{1\perp}\max\{1,\zeta\})
- \Theta\left(p_{T}^{\rm
                                      veto}-k_{1\perp}\sqrt{1+\zeta^2+2\zeta\cos\phi}\right)\right]\\
                                & \times \Theta(\eta^2+\phi^2 - R^2)\,,
\end{split}
\end{align}
where we used the explicit form of $\Theta_{\rm cluster}(R^2)$ in the
variables defined above, namely the relation
\begin{align}\begin{split}
  \Theta(p_{T}^{\rm veto}-\max\{p_T^{\rm jet_i}\})
\Theta_{\rm cluster}(R^2)\equiv
\Theta(p_{T}^{\rm veto}-k_{1\perp}\max\{1,\zeta\})\Theta(\eta^2+\phi^2 - R^2)\\
+\Theta\left(p_{T}^{\rm veto}-k_{1\perp}\sqrt{1+\zeta^2+2\zeta\cos\phi}\right)
\Theta(R^2-\eta^2-\phi^2)\,,
\end{split}\end{align}
followed by
\begin{align} 
    \Theta(R^2-\eta^2-\phi^2) = 1 - \Theta(\eta^2+\phi^2 - R^2) \,.
\end{align}

With the above parametrisation, the correlated and uncorrelated
contributions to the squared amplitude factorise as (see
e.g. Refs.~\cite{Banfi:2012yh,Banfi:2014sua})
\begin{align}
{\cal A}^{\rm cor./uncor.}(k_1,k_2) &\equiv
\frac{1}{k_{1\perp}^4}\frac{1}{\zeta^2} {\cal D}^{\rm cor./uncor.}(\zeta,\eta,\phi)\,,
\end{align}
where the function ${\cal D}^{\rm cor./uncor.}(\zeta,\eta,\phi)$ is
regular in the limit $\zeta \to 0$. In particular one has
\begin{equation}
 \label{eq:Duncor}
  {\cal D}^{\rm uncor.}(\zeta,\eta,\phi)=16\,C_R^2\,.
\end{equation}
The calculation of $\Delta{\cal S}^{(2)}(p_{T},R^2;\mu,\nu)$ will then
involve the evaluation of phase-space integrals of the type
\begin{equation}
 \label{eq:MIs}
\int \frac{ \rd k_{1\perp}}{k_{1\perp}} \,\rd\eta_1 \,\frac{\rd\zeta}{\zeta}\,\rd\eta \,\frac{\rd \phi}{2\pi}\, e^{-2 k_{1\perp} \frac{e^{-\gamma_E}}{\nu}[\cosh{(\eta_1)}+\zeta\cosh{(\eta-\eta_1)}]}\,{\cal D}(\zeta,\eta,\phi)\, \Delta {\cal M}(p_{T}^{\rm veto},R^2)\,,
\end{equation}
with $\Delta {\cal M}(p_{T}^{\rm veto},R^2)$ being defined in
Eq.~\eqref{eq:DeltaJ-2L}, and where we have already performed the
integration over $\phi_1$.
We now discuss separately the correlated and uncorrelated
contributions.
\paragraph{The correlated correction.}
We start with the correlated corrections. The function
${\cal D}^{\rm cor.}(\zeta,\eta,\phi)$ vanishes in the limits
$\eta \to \pm \infty$ by definition of this contribution, and
therefore the only source of rapidity divergences in
Eq.~\eqref{eq:MIs} are the limits $\eta_1\to \pm \infty$.
We can then integrate over $\eta_1$ and expand the result around
$\nu\to \infty$ and, according to our regularisation
procedure, retain only the leading term. We write
\begin{align} \begin{split}\label{eq:flatetaint}
\int_{-\infty}^{+\infty} \rd\eta_1 & e^{-2 k_{1\perp}
  \frac{e^{-\gamma_E}}{\nu}[\cosh{(\eta_1)}+\zeta\cosh{(\eta-\eta_1)}]}
                                      = \Omega\left(\frac{k_{1\perp}}{\nu},\zeta,\eta\right)  + {\cal O}\left(\frac{1}{\nu^2}\right)\,,\\
&\Omega\left(\frac{k_{1\perp}}{\nu},\zeta,\eta\right)\equiv \eta + 2\ln \frac{\nu}{k_{1\perp}} - \ln \left(1 + \zeta
  e^\eta\right) - \ln \left(\zeta+
  e^\eta\right)\,,
\end{split}\end{align}
so that Eq.~\eqref{eq:MIs} leads to the integral
\begin{align}\label{eq:MIs-cor}
I\left(\frac{p_{T}^{\rm veto}}{\nu},R^2\right)=\int \frac{ \rd
  k_{1\perp}}{k_{1\perp}} 
  \,\frac{\rd\zeta}{\zeta}\,\rd\eta \,\frac{\rd \phi}{2\pi}\,
  \Omega\left(\frac{k_{1\perp}}{\nu},\zeta,\eta\right) \,{\cal D}^{\rm
  cor.}(\zeta,\eta,\phi)\, \Delta {\cal M}(p_{T}^{\rm veto},R^2)\,.
\end{align}

We then decompose the $\Theta(\eta^2+\phi^2 - R^2)$ function in
$\Delta {\cal M}(p_{T}^{\rm veto},R^2)$ given in
Eq.~\eqref{eq:DeltaJ-2L} as
\begin{align}\label{eq:decomposition}
\Theta \left(\eta ^2-R^2+\phi ^2\right)=\underbrace{\Theta \left(\phi ^2-R^2\right)}_{{\rm part\;}  A}+
\underbrace{\Theta \left(R^2-\phi ^2\right) \Theta \left(\eta ^2-R^2+\phi ^2\right)}_{{\rm part\;} B}\,.
\end{align}
The integral $I_A$ associated with part $A$ is simple to evaluate by
direct integration using the package
\texttt{PolyLogTools}~\cite{Duhr:2019tlz}. We obtained the exact $R^2$
dependence and subsequently expanded the final result in a Laurent
series about $R^2=0$.

To compute the integral corresponding to part $B$, we consider its
derivative with respect to the (squared) jet radius $R^2$.
Equation~\eqref{eq:MIs-cor} leads to the differential equation
\begin{align}\label{eq:MIs-cor-de}
\frac{\partial }{\partial R^2}I_B\left(\frac{p_{T}^{\rm veto}}{\nu},R^2\right)&=\int \frac{ \rd
  k_{1\perp}}{k_{1\perp}} 
  \,\frac{\rd\zeta}{\zeta}\,\rd\eta \,\frac{\rd \phi}{2\pi}\,
  \Omega\left(\frac{k_{1\perp}}{\nu},\zeta,\eta\right) \,{\cal D}^{\rm
  cor.}(\zeta,\eta,\phi)\, \\
  &\times \left[\Theta(p_{T}^{\rm veto}-k_{1\perp}\max\{1,\zeta\})
- \Theta\left(p_{T}^{\rm
    veto}-k_{1\perp}\sqrt{1+\zeta^2+2\zeta\cos\phi}\right)\right]\notag\\
  & \times \left(\delta(R^2 - \phi^2) 
-  \Theta \left(R^2-\phi ^2\right)\delta \left(\eta ^2-R^2+\phi ^2\right)\right)\,,\notag
\end{align}
We compute the right-hand side of the above equation as a Laurent
series in $R^2$, up to and including $R^8$ terms.
As a boundary condition, we take the limit of $I_B$ for $R=0$, which
is allowed by the fact that the entire collinear divergence in
Eq.~\eqref{eq:decomposition} is encoded in part $A$.  Some care is
needed in applying the method of expansion by
regions~\cite{Beneke:1997zp} when taking this limit. Starting from
${\cal O}(R^2)$ more than one region contributes which requires the
introduction of additional regulators. The calculation of the
${\cal O}(1)$ term, however, only receives contributions from a single
region and the integral is well defined.

We apply the above strategy to both colour structures
contributing to $\cor(p_{T},R^2;\mu,\nu)$. The resulting
expressions are provided in electronic form in the ancillary files
\texttt{SoftFunctionFermion.m} and 
\texttt{SoftFunctionNonAbelian.m}
for $C_R n_F T_R$ and $C_RC_A$, respectively.\sloppy

\paragraph{The uncorrelated correction.}
We now consider the uncorrelated case. Thanks to Eq.~\eqref{eq:Duncor},
the integrand takes a simpler form than in the correlated case:
\begin{equation}
 \label{eq:MIs-uncor-1}
\int \frac{ \rd k_{1\perp}}{k_{1\perp}} \,\rd\eta_1 \,\frac{\rd\zeta}{\zeta}\,\rd\eta \,\frac{\rd \phi}{2\pi}\, e^{-2 k_{1\perp} \frac{e^{-\gamma_E}}{\nu}[\cosh{(\eta_1)}+\zeta\cosh{(\eta-\eta_1)}]}\,\, \Delta {\cal M}(p_{T}^{\rm veto},R^2)\,.
\end{equation}
Despite its apparent simplicity, this integral has a more complicated 
structure of rapidity divergences, which now
originate both from the $\eta_1\to\pm\infty$ and $\eta\to\pm\infty$
limits.
To handle this situation, we start by taking a Laplace transform of
the exponential regulator with respect to the variable
$e^{-\gamma_E}/\nu$ and denote its conjugate by $\tau$. 
Introducing the more convenient variables
\begin{align}
w = e^{\eta}, \quad x = e^{\eta_1}, \label{eq:var_change}
\end{align} 
the Laplace transform gives (including the Jacobian corresponding to the above change of
variable)
\begin{align}
\label{eq:laplace}
\frac{1}{x\,w}e^{-k_{1\perp}
  \frac{e^{-\gamma_E}}{\nu\,x}[1+w\zeta+\frac{x^2}{w}(w+\zeta)]}\,\LTeq\,
  \frac{1}{ k_{1\perp}\left(w+w x^2 + w^2 \zeta + x^2\zeta\right) + w\,
  x\, \tau}\,.
\end{align}
We then perform an expansion of the exponential regulator in
distributions in both $x$ and $w$ in Laplace space, where the
exponential regulator becomes the rational function in Eq.~\eqref{eq:laplace}. 
Finally, we take the leading term in the limit of $\tau \to \infty$
(corresponding to $\nu \to \infty$) and take the inverse Laplace
transform of the result. This procedure yields
\begin{align}
\int\frac{\rd x}{x\,w}e^{-k_{1\perp}
  \frac{e^{-\gamma_E}}{\nu\,x}[1+w\zeta+\frac{x^2}{w}(w+\zeta)]}
  &\to 4  \delta (w) \ln \left(\frac{k_{1 \perp }}{\nu }\right) \ln
    \left(\frac{\zeta\,k_{1\perp}}{\nu }\right)\notag\\
  &\quad\quad+
  \left[\frac{1}{w}\right]_+\,\ln \left(\frac{\nu^2 w  }{k_{1\perp}^2
    (w+\zeta)(1+\zeta w)}\right) + {\cal O}\left(\frac{1}{\nu^2}\right)\,,
\end{align}
where we also evaluated the $x$ integral using the fact that the
remaining factors in the integrand of Eq.~\eqref{eq:MIs-uncor-1} are
$x$ independent.
For the remaining integrals, we follow the same procedure as for the
calculation of the correlated term $\cor(p_{T},R^2;\mu,\nu)$. We
report the resulting expression in the ancillary file
\texttt{SoftFunctionAbelian.m}.

\subsubsection{Numerical computation of the two-loop soft function}
\label{sec:twoloop-num}

In this section we briefly outline the procedure used for the
numerical evaluation of the quantity $\Delta \mathcal{S}^{(2)}$,
retaining the full dependence on the jet radius. 
This independent calculation will provide a non-trivial check
of the analytic calculation discussed in the previous section.
Numerical calculations were performed to a precision at the permille
level.

\paragraph{The correlated correction.} We express the integral for
$\Delta \mathcal{S}^{(2)}$ in terms of the variables
\begin{align}
  \eta_t = \frac{1}{2}\left( \eta_1 + \eta_2\right) \,,\qquad \eta = \eta_1 - \eta_2 \,,\qquad
  \mathcal{K}^2_T = k_{1\perp}^2 + k_{2\perp}^2\,,  \qquad z = \dfrac{k_{1\perp}^2}{k_{1\perp}^2 + k_{2\perp}^2}\,,
\end{align}
in addition to $\phi$, as defined in Eq.~\eqref{eq:coordinates}.

For the correlated contribution, the integrand is strongly peaked around 
$\eta=0$, meaning that no rapidity regulator is needed for this integration.
The integrand is however flat in $\eta_t$, 
leading to a rapidity divergence when integrating over this variable. 
The structure of the integral over $\eta_t$, including the exponential
regulator factor, is exactly that of Eq.~\eqref{eq:flatetaint}, and we
use this equation to perform the integral analytically (in the
$\nu \to \infty$ limit). The structure of the resulting integrand in
the variable $\mathcal{K}^2_T$ is very simple, containing only terms
of the form $\ln^n\left(\mathcal{K}^2_T/\nu^2\right)/\mathcal{K}^2_T$,
and we can also perform the integration over $\mathcal{K}^2_T$
analytically. Finally, we are left with (finite) integrations over the
$\eta$, $\phi$, and $z$ variables. These are performed numerically
using the \texttt{GlobalAdaptive} \texttt{NIntegrate} routine from
\texttt{Mathematica}.

\paragraph{The uncorrelated correction.} Here we re-express the
constraint in the measurement function for $\Delta \mathcal{S}^{(2)}$
as
\begin{align} \label{eq:delRsimpleflip}
    \Theta(\eta^2+\phi^2 - R^2) = 1 - \Theta(R^2-\eta^2-\phi^2)\,.
\end{align}

For the second term on the right-hand side, there is guaranteed to be
no rapidity divergence in the integration over $\eta$ by definition,
and we may use exactly the same techniques as were used for the
correlated correction. For the first term, we use the integration
variables $\eta_1$, $\eta_2$, $\mathcal{K}_T^2$, $z$ and $\phi$. The
integrand does not depend on either $\eta_1$ or $\eta_2$, such that we
experience rapidity divergences in the integrations over both of these
variables. After inserting the exponential regulator, we may perform
the integrations over both $\eta_1$ and $\eta_2$ analytically by
making use of the result in Eq.~\eqref{eq:flatetaint} (where here we
only need the result for the case where $\zeta=0$). As for the
correlated case, the structure of the integral in $\mathcal{K}^2_T$ is
simple and may also be performed analytically. This then leaves us
with the $z$ and $\phi$ integrations, which are performed numerically
via the \texttt{GlobalAdaptive} \texttt{NIntegrate} routine from
\texttt{Mathematica}.


\section{Results and dependence on the jet radius}
\label{sec:results}
We now discuss some consistency checks of our results and the validity
of the small-$R$ expansion for phenomenologically relevant values of
the jet radius.

We start by extracting the rapidity anomalous dimensions, which can be
compared with the results of
Refs.~\cite{Banfi:2012jm,Becher:2013xia,Stewart:2013faa}. Specifically,
starting from the results for
$\Delta{\cal S}^{(2)}(p_{T},R^2;\mu,\nu)$ computed in the previous
section we can obtain the \textit{difference} between the rapidity
anomalous dimension entering the resummation of $p^{\rm veto}_T$ and
that of $p_T$.
At $\mathcal{O}(\alpha_s^2)$ this amounts to computing:
\begin{align}\begin{split}
 \label{eq:gammanucfnf}
&\frac{\partial \Delta{\cal S}^{(2)}_{C_R \,n_F \,T_R}(p_{T},R^2;\mu,\nu)}{\partial
  \ln\nu} 
  = C_R \,n_F \,T_R\,\Bigg[\frac{16}{9} (24 \ln (2)-23) \ln
  (R)\\
  &\quad -\frac{16}{27} \left(-157+6 \pi ^2+72 \ln ^2(2)+96 \ln
  (2)\right) + \frac{3071-1680 \,\ln (2)}{1350} R^2+ {\cal O}\left(R^4\right)\Bigg]\,,
\end{split}\end{align}
\begin{align}\begin{split}
 \label{eq:gammanucfca}
 \frac{\partial \Delta{\cal S}^{(2)}_{C_R C_A}(p_{T},R^2;\mu,\nu)}{\partial
  \ln\nu} 
  =\,& C_R C_A\,\Bigg[-\frac{8}{9} \left(-131+12
                          \,\pi ^2+132\, \ln (2)\right) \ln (R)\\
 &  +\frac{8}{27} \left(108 \zeta_3-805+33 \,\pi ^2+396
   \ln ^2(2)+420\, \ln (2)\right) \\
  &  + \frac{1429+3600 \,\pi
                          ^2+12480\, \ln (2)}{2700}  R^2 + {\cal O}\left(R^4\right)\Bigg]\,,
\end{split}\end{align}
\begin{align}
 \label{eq:gammanucf2}
& \frac{\partial \Delta{\cal S}^{(2)}_{C_R^2}(p_{T},R^2;\mu,\nu)}{\partial
  \ln\nu} 
  = C_R^2\,\Bigg[64 \,\zeta_3 -\frac{16 }{3}\,\pi ^2 R^2+4 \,R^4\Bigg]\,,
\end{align}
where $C_R$ is the quadratic Casimir in the representation of the
initial-state partons (either quarks or gluons). 
Equations \eqref{eq:gammanucfnf} and \eqref{eq:gammanucfca} are
obtained from the correlated corrections, while
Eq.~\eqref{eq:gammanucf2} is obtained from the uncorrelated
corrections.
For the $C_R^2$ colour factor we present the full $R^2$ dependence,
since the series terminates at order $R^4$ for the anomalous dimension
if $R<\pi$ (see the discussion in the appendix of
Ref.~\cite{Banfi:2012yh} for $R > \pi$).  For the other two
colour factors, higher-order terms in $R^2$ are not given here for
simplicity but can be obtained from the results in the ancillary
files.
These results reproduce those given in
Refs.~\cite{Banfi:2012jm,Becher:2013xia,Stewart:2013faa}.
A numerical calculation of the same soft function defined in
Eq.~\eqref{eq:soft_def} has been previously presented in
Ref.~\cite{Bell:2020yzz} using the \texttt{SoftServe}
code~\cite{Bell:2018vaa}.
This computation uses a different rapidity regularisation procedure to
ours. As a consequence, while we find agreement at the level of the
anomalous dimensions, the boundary conditions to
Eqs.~\eqref{eq:soft_RGES} are scheme dependent and thus differ from
the result of Ref.~\cite{Bell:2020yzz}.  \sloppy

\begin{figure}[t!]
  \centering
  \includegraphics[width=0.6\linewidth]{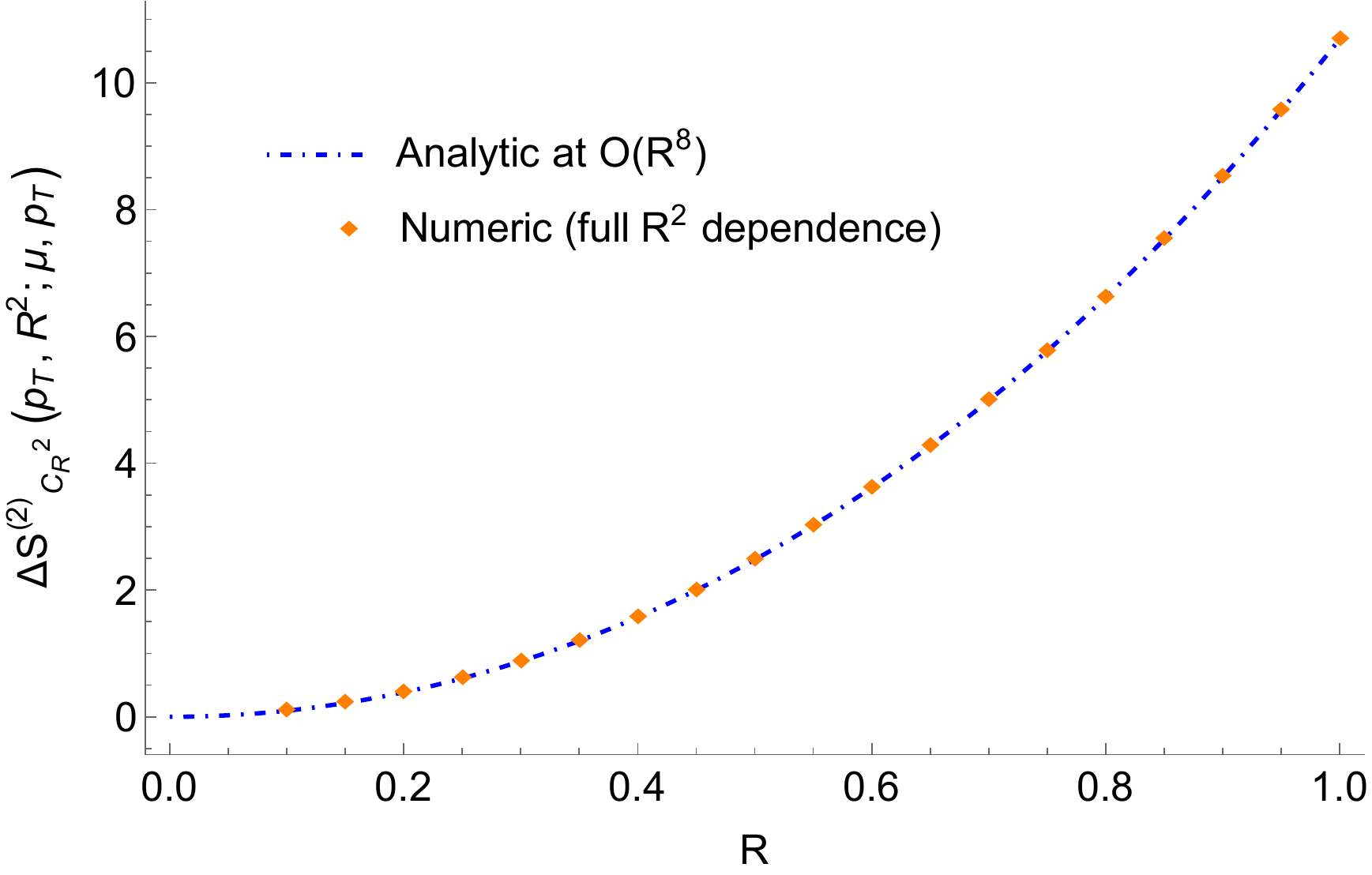}
  \includegraphics[width=0.6\linewidth]{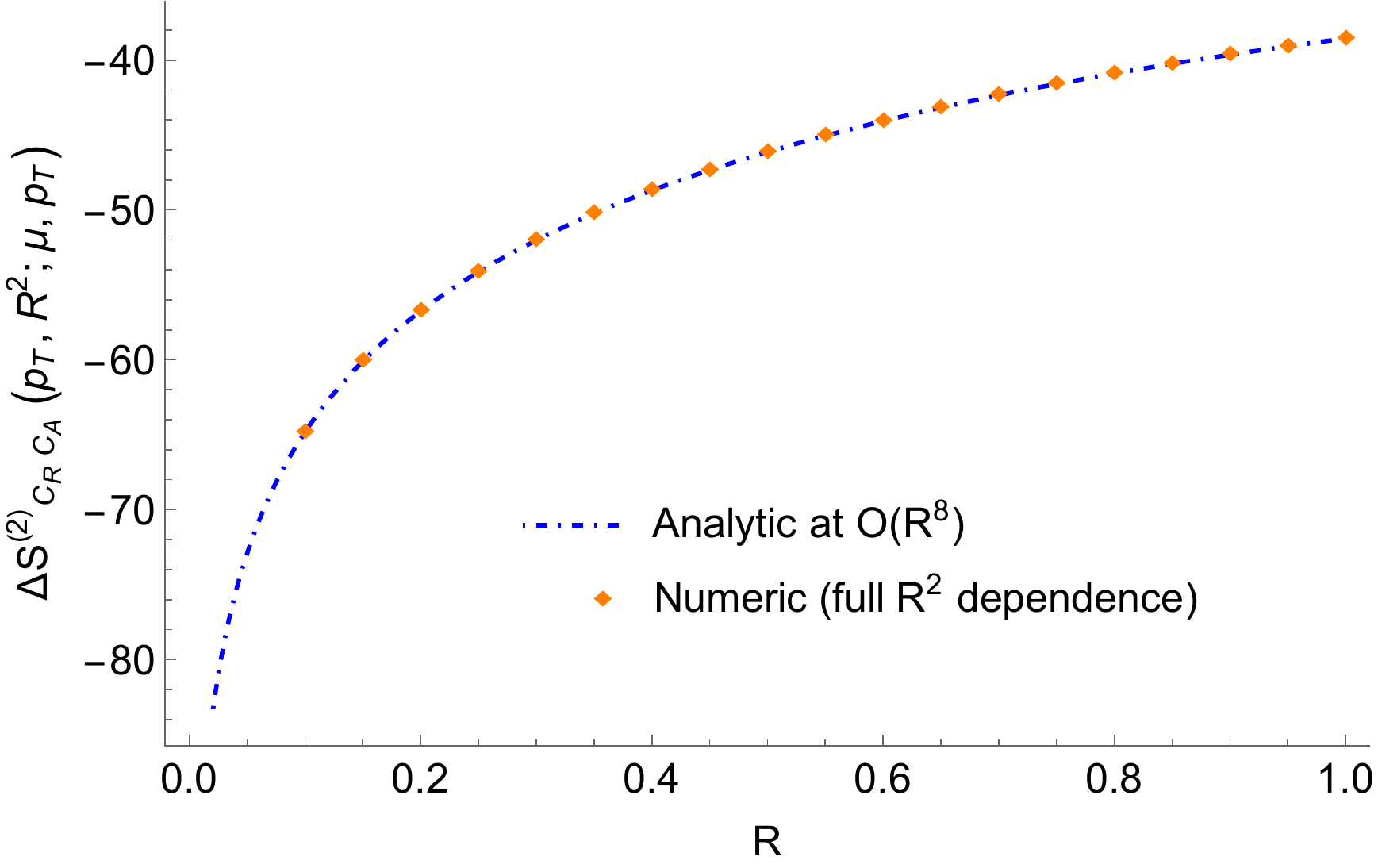}
  \includegraphics[width=0.6\linewidth]{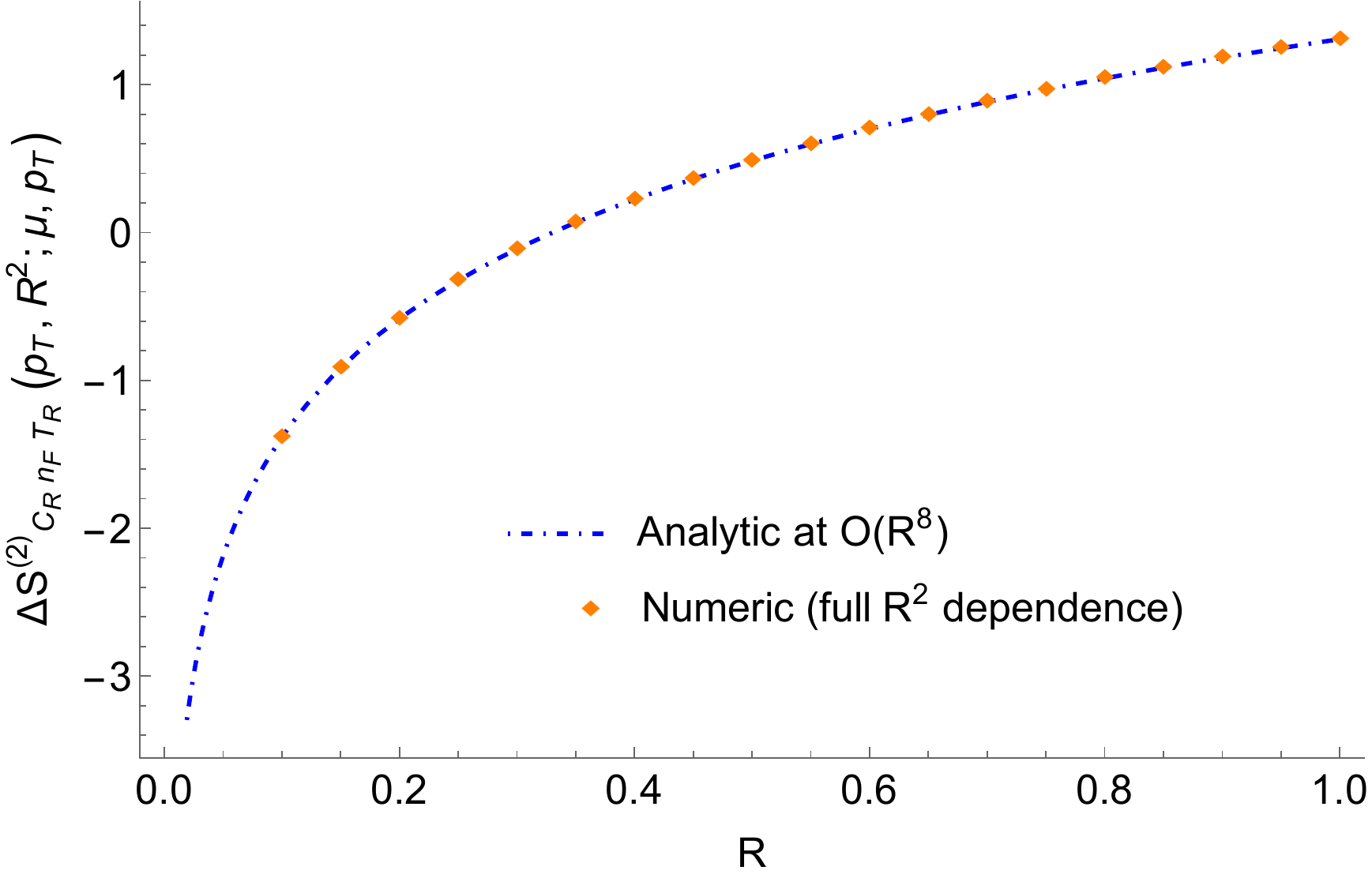}
  \caption{\label{fig:Rdep} Comparison between the numerical
    calculation with full $R^2$ dependence (orange diamonds) and the analytic
    expansion through ${\cal O}(R^8)$ (blue dot-dashed lines) for the three
    colour structures contributing to the two-loop soft function.}
\end{figure}

\begin{figure}[t!]
  \centering
  \includegraphics[width=0.6\linewidth]{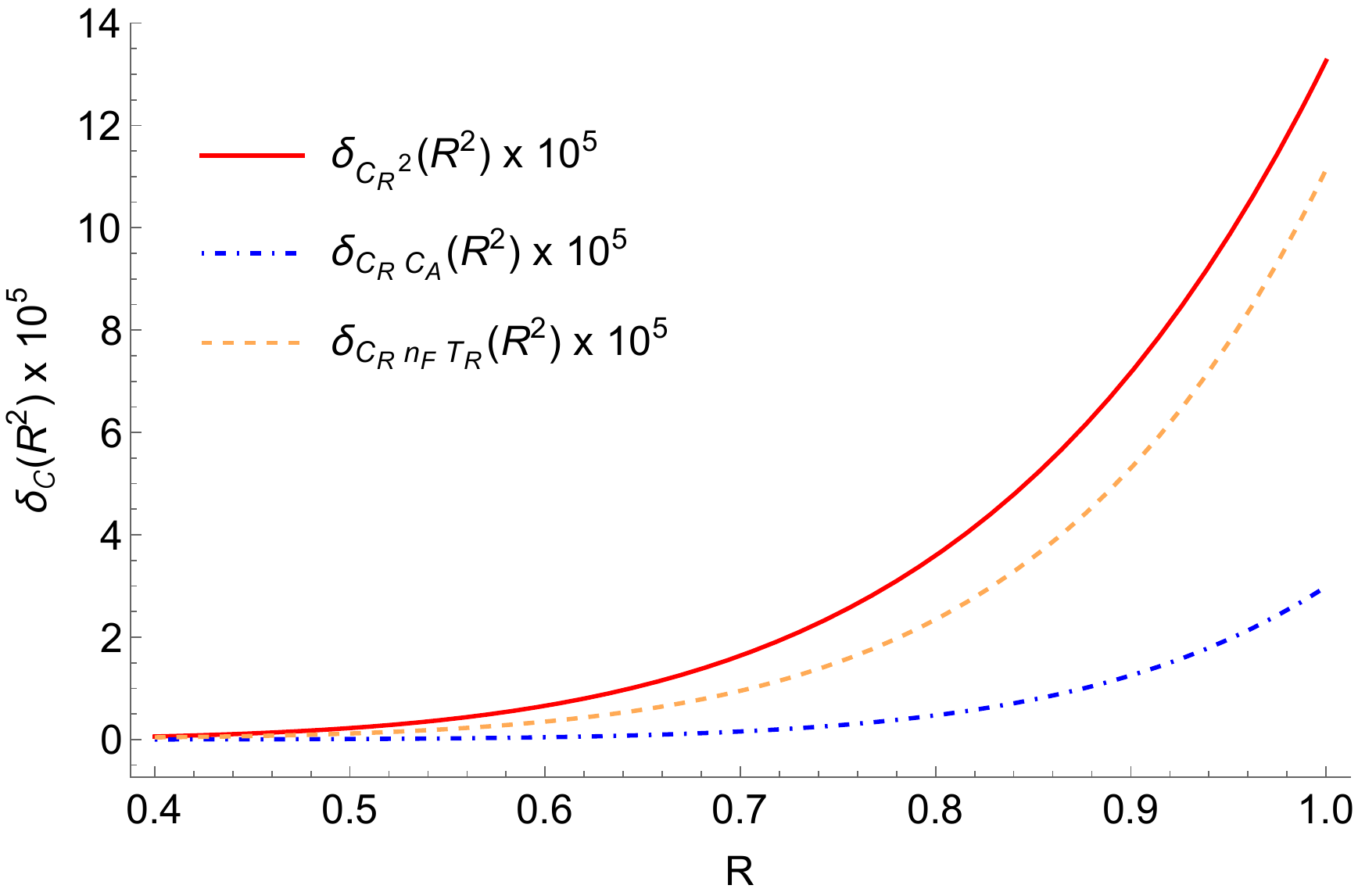}
  \caption{\label{fig:smallR} The quantity $\delta_{\mathcal{C}}(R^2)$
    defined in Eq.~\eqref{eq:delta} for the three colour structures
    contributing to the two-loop soft function. The three curves are
    multiplied by a factor of $10^5$.}
\end{figure}

We next assess the validity of the small-$R$ expansion used in the
calculation of the soft function, and specifically whether this
expansion is sufficiently accurate for typical values of the jet
radius $R\in [0.2,0.8]$.  We first compare the analytic expansion in
$R^2$ to the numerical calculation of the soft function, whose $R^2$
dependence is exact. Fig.~\ref{fig:Rdep} displays the $R$ dependence
of the two results for $\nu=p_T$, 
which agree well in the considered range of~$R$.
As a second check, we consider the quantity 
$\Delta{\cal S}^{(2)}(p_{T},R^2;\mu,\nu)$ truncated at
different orders in $R^2$ for the three different colour
structures. More precisely, we define the relative difference of the
expansions at sixth and eighth order in $R$ (for $\nu=p_{T}$), and
plot the quantity
\begin{equation}\label{eq:delta}
  \delta_{\mathcal{C}}(R)=\left|1-\frac{\Delta{\cal S}^{(2)}_\mathcal{C} (p_{T},R^2;\mu, p_{T})|_{R^6}}
  {\Delta{\cal S}^{(2)}_{\mathcal{C}} (p_{T},R^2;\mu, p_{T}) |_{R^8}}\right|\,,
\end{equation}
for the three different colour factors, $\mathcal{C}\in\{C_R n_F T_R, C_R C_A,C^2_R\}$
in Fig.~\ref{fig:smallR}.
The plot shows an excellent convergence of the small-$R$ expansion all the way
up to $R=1.0$, with residual corrections being at the sub-permille
level.
As a benchmark, in Table \ref{tab:convergence} we present the values
of $\delta_{\mathcal{C}}(R^2)$ for $R=0.4$ and $R=0.8$, showing that
${\cal O}(R^8)$ terms are indeed numerically negligible.
The two figures~\ref{fig:Rdep}--\ref{fig:smallR} indicate that the
truncation error associated with our small-$R$ expansion is well under
control for phenomenologically relevant values of the jet radius.
We note nevertheless that the same strategy that was used here to
obtain the results up to ${\cal O}(R^8)$ allows us to compute
higher-order terms should they be required.
\begin{table}
\centering
\begin{tabular}{c | c c c} 
   & $C_R n_F T_R$ & $C_R C_A$ & $C_R^2$ \\ [0.5ex] 
 \hline
 $R=0.4$ & $4.3\times10^{-7}$ & $1.6\times 10^{-8}$ & $5.9\times 10^{-7}$ \\ 
 \hline
 $R=0.8$ & $2.3\times 10^{-5}$ & $4.7\times 10^{-6}$ & $3.6\times 10^{-5}$ 
\end{tabular}
\caption{Reference values for $\delta_{\mathcal{C}}(R^2)$, as defined in Eq.~\eqref{eq:delta}.}
\label{tab:convergence}
\end{table}

\section{Conclusions}
\label{sec:conclusions}
In this article, we presented the first complete analytic calculation
of the two-loop soft function which enters the factorisation theorem
for the leading-jet transverse momentum resummation, or equivalently
the jet-vetoed cross section for the production of any colour singlet
system.
We carry out two independent calculations: an analytic expansion for
small values of the jet radius $R$ up to and including ${\cal O}(R^8)$
terms; and a numerical calculation for selected values of $R$ where
the full dependence on $R$ is retained.
The two calculations agree perfectly for selected $R$ values in the
range $R \in [0,1]$, relevant to applications to collider
phenomenology.
All results are attached to the \texttt{arXiv} submission of this
article in \texttt{Mathematica} readable files.
This work constitutes a first important step towards the N$^3$LL
resummation of the leading-jet transverse momentum distribution.
Among the missing ingredients which are currently unknown, one needs
the two-loop beam functions as well as the three-loop rapidity
anomalous dimension.
Moreover, going beyond NNLL requires also a careful formulation of the
factorisation theorem in SCET.
We will address the above points in forthcoming publications.

\acknowledgments
We are grateful to Luca Rottoli for discussions.
The work of JRG is supported by the Royal Society through Grant
URF\textbackslash R1\textbackslash 201500.
RS is supported by the United States Department of Energy under Grant
Contract DE-SC0012704.

\bibliographystyle{JHEP} \bibliography{soft}

\providecommand{\href}[2]{#2}\begingroup\raggedright\begin{thebibliography}{10}

\bibitem{Banfi:2012yh}
A.~Banfi, G.~P. Salam and G.~Zanderighi, \emph{{NLL+NNLO predictions for
  jet-veto efficiencies in Higgs-boson and Drell-Yan production}},
  \href{http://dx.doi.org/10.1007/JHEP06(2012)159}{\emph{JHEP} {\bf 06} (2012)
  159}, [\href{http://arxiv.org/abs/1203.5773}{{\tt 1203.5773}}].

\bibitem{Banfi:2012jm}
A.~Banfi, P.~F. Monni, G.~P. Salam and G.~Zanderighi, \emph{{Higgs and Z-boson
  production with a jet veto}},
  \href{http://dx.doi.org/10.1103/PhysRevLett.109.202001}{\emph{Phys. Rev.
  Lett.} {\bf 109} (2012) 202001}, [\href{http://arxiv.org/abs/1206.4998}{{\tt
  1206.4998}}].

\bibitem{Becher:2013xia}
T.~Becher, M.~Neubert and L.~Rothen, \emph{{Factorization and
  $N^{3}LL_{p}$+NNLO predictions for the Higgs cross section with a jet veto}},
  \href{http://dx.doi.org/10.1007/JHEP10(2013)125}{\emph{JHEP} {\bf 10} (2013)
  125}, [\href{http://arxiv.org/abs/1307.0025}{{\tt 1307.0025}}].

\bibitem{Stewart:2013faa}
I.~W. Stewart, F.~J. Tackmann, J.~R. Walsh and S.~Zuberi, \emph{{Jet $p_T$
  resummation in Higgs production at $NNLL'+NNLO$}},
  \href{http://dx.doi.org/10.1103/PhysRevD.89.054001}{\emph{Phys. Rev. D} {\bf
  89} (2014) 054001}, [\href{http://arxiv.org/abs/1307.1808}{{\tt 1307.1808}}].

\bibitem{Bauer:2000yr}
C.~W. Bauer, S.~Fleming, D.~Pirjol and I.~W. Stewart, \emph{{An Effective field
  theory for collinear and soft gluons: Heavy to light decays}},
  \href{http://dx.doi.org/10.1103/PhysRevD.63.114020}{\emph{Phys. Rev. D} {\bf
  63} (2001) 114020}, [\href{http://arxiv.org/abs/hep-ph/0011336}{{\tt
  hep-ph/0011336}}].

\bibitem{Bauer:2001yt}
C.~W. Bauer, D.~Pirjol and I.~W. Stewart, \emph{{Soft collinear factorization
  in effective field theory}},
  \href{http://dx.doi.org/10.1103/PhysRevD.65.054022}{\emph{Phys. Rev. D} {\bf
  65} (2002) 054022}, [\href{http://arxiv.org/abs/hep-ph/0109045}{{\tt
  hep-ph/0109045}}].

\bibitem{Bauer:2002nz}
C.~W. Bauer, S.~Fleming, D.~Pirjol, I.~Z. Rothstein and I.~W. Stewart,
  \emph{{Hard scattering factorization from effective field theory}},
  \href{http://dx.doi.org/10.1103/PhysRevD.66.014017}{\emph{Phys. Rev. D} {\bf
  66} (2002) 014017}, [\href{http://arxiv.org/abs/hep-ph/0202088}{{\tt
  hep-ph/0202088}}].

\bibitem{Beneke:2002ph}
M.~Beneke, A.~P. Chapovsky, M.~Diehl and T.~Feldmann, \emph{{Soft collinear
  effective theory and heavy to light currents beyond leading power}},
  \href{http://dx.doi.org/10.1016/S0550-3213(02)00687-9}{\emph{Nucl. Phys. B}
  {\bf 643} (2002) 431--476}, [\href{http://arxiv.org/abs/hep-ph/0206152}{{\tt
  hep-ph/0206152}}].

\bibitem{Beneke:2002ni}
M.~Beneke and T.~Feldmann, \emph{{Multipole expanded soft collinear effective
  theory with nonAbelian gauge symmetry}},
  \href{http://dx.doi.org/10.1016/S0370-2693(02)03204-5}{\emph{Phys. Lett. B}
  {\bf 553} (2003) 267--276}, [\href{http://arxiv.org/abs/hep-ph/0211358}{{\tt
  hep-ph/0211358}}].

\bibitem{Banfi:2015pju}
A.~Banfi, F.~Caola, F.~A. Dreyer, P.~F. Monni, G.~P. Salam, G.~Zanderighi
  et~al., \emph{{Jet-vetoed Higgs cross section in gluon fusion at
  N$^{3}$LO+NNLL with small-$R$ resummation}},
  \href{http://dx.doi.org/10.1007/JHEP04(2016)049}{\emph{JHEP} {\bf 04} (2016)
  049}, [\href{http://arxiv.org/abs/1511.02886}{{\tt 1511.02886}}].

\bibitem{Banfi:2013eda}
A.~Banfi, P.~F. Monni and G.~Zanderighi, \emph{{Quark masses in Higgs
  production with a jet veto}},
  \href{http://dx.doi.org/10.1007/JHEP01(2014)097}{\emph{JHEP} {\bf 01} (2014)
  097}, [\href{http://arxiv.org/abs/1308.4634}{{\tt 1308.4634}}].

\bibitem{Dasgupta:2014yra}
M.~Dasgupta, F.~Dreyer, G.~P. Salam and G.~Soyez, \emph{{Small-radius jets to
  all orders in QCD}},
  \href{http://dx.doi.org/10.1007/JHEP04(2015)039}{\emph{JHEP} {\bf 04} (2015)
  039}, [\href{http://arxiv.org/abs/1411.5182}{{\tt 1411.5182}}].

\bibitem{Liu:2012sz}
X.~Liu and F.~Petriello, \emph{{Resummation of jet-veto logarithms in hadronic
  processes containing jets}},
  \href{http://dx.doi.org/10.1103/PhysRevD.87.014018}{\emph{Phys. Rev. D} {\bf
  87} (2013) 014018}, [\href{http://arxiv.org/abs/1210.1906}{{\tt 1210.1906}}].

\bibitem{Boughezal:2013oha}
R.~Boughezal, X.~Liu, F.~Petriello, F.~J. Tackmann and J.~R. Walsh,
  \emph{{Combining Resummed Higgs Predictions Across Jet Bins}},
  \href{http://dx.doi.org/10.1103/PhysRevD.89.074044}{\emph{Phys. Rev. D} {\bf
  89} (2014) 074044}, [\href{http://arxiv.org/abs/1312.4535}{{\tt 1312.4535}}].

\bibitem{Monni:2019yyr}
P.~F. Monni, L.~Rottoli and P.~Torrielli, \emph{{Higgs transverse momentum with
  a jet veto: a double-differential resummation}},
  \href{http://dx.doi.org/10.1103/PhysRevLett.124.252001}{\emph{Phys. Rev.
  Lett.} {\bf 124} (2020) 252001}, [\href{http://arxiv.org/abs/1909.04704}{{\tt
  1909.04704}}].

\bibitem{Michel:2018hui}
J.~K.~L. Michel, P.~Pietrulewicz and F.~J. Tackmann, \emph{{Jet Veto
  Resummation with Jet Rapidity Cuts}},
  \href{http://dx.doi.org/10.1007/JHEP04(2019)142}{\emph{JHEP} {\bf 04} (2019)
  142}, [\href{http://arxiv.org/abs/1810.12911}{{\tt 1810.12911}}].

\bibitem{Gangal:2014qda}
S.~Gangal, M.~Stahlhofen and F.~J. Tackmann, \emph{{Rapidity-Dependent Jet
  Vetoes}}, \href{http://dx.doi.org/10.1103/PhysRevD.91.054023}{\emph{Phys.
  Rev. D} {\bf 91} (2015) 054023}, [\href{http://arxiv.org/abs/1412.4792}{{\tt
  1412.4792}}].

\bibitem{Gangal:2016kuo}
S.~Gangal, J.~R. Gaunt, M.~Stahlhofen and F.~J. Tackmann, \emph{{Two-Loop Beam
  and Soft Functions for Rapidity-Dependent Jet Vetoes}},
  \href{http://dx.doi.org/10.1007/JHEP02(2017)026}{\emph{JHEP} {\bf 02} (2017)
  026}, [\href{http://arxiv.org/abs/1608.01999}{{\tt 1608.01999}}].

\bibitem{Gangal:2020qik}
S.~Gangal, J.~R. Gaunt, F.~J. Tackmann and E.~Vryonidou, \emph{{Higgs
  Production at NNLL$'$+NNLO using Rapidity Dependent Jet Vetoes}},
  \href{http://dx.doi.org/10.1007/JHEP05(2020)054}{\emph{JHEP} {\bf 05} (2020)
  054}, [\href{http://arxiv.org/abs/2003.04323}{{\tt 2003.04323}}].

\bibitem{Becher:2014aya}
T.~Becher, R.~Frederix, M.~Neubert and L.~Rothen, \emph{{Automated NNLL $+$ NLO
  resummation for jet-veto cross sections}},
  \href{http://dx.doi.org/10.1140/epjc/s10052-015-3368-y}{\emph{Eur. Phys. J.
  C} {\bf 75} (2015) 154}, [\href{http://arxiv.org/abs/1412.8408}{{\tt
  1412.8408}}].

\bibitem{Moult:2014pja}
I.~Moult and I.~W. Stewart, \emph{{Jet Vetoes interfering with $H \to WW$}},
  \href{http://dx.doi.org/10.1007/JHEP09(2014)129}{\emph{JHEP} {\bf 09} (2014)
  129}, [\href{http://arxiv.org/abs/1405.5534}{{\tt 1405.5534}}].

\bibitem{Jaiswal:2014yba}
P.~Jaiswal and T.~Okui, \emph{{Explanation of the $WW$ excess at the LHC by
  jet-veto resummation}},
  \href{http://dx.doi.org/10.1103/PhysRevD.90.073009}{\emph{Phys. Rev. D} {\bf
  90} (2014) 073009}, [\href{http://arxiv.org/abs/1407.4537}{{\tt 1407.4537}}].

\bibitem{Monni:2014zra}
P.~F. Monni and G.~Zanderighi, \emph{{On the excess in the inclusive $
  {W}^{+}{W}^{-}\ \to\ {l}^{+}{l}^{-}\nu \overline{\nu} $ cross section}},
  \href{http://dx.doi.org/10.1007/JHEP05(2015)013}{\emph{JHEP} {\bf 05} (2015)
  013}, [\href{http://arxiv.org/abs/1410.4745}{{\tt 1410.4745}}].

\bibitem{Dawson:2016ysj}
S.~Dawson, P.~Jaiswal, Y.~Li, H.~Ramani and M.~Zeng, \emph{{Resummation of jet
  veto logarithms at N$^3$LL$_a$ + NNLO for $W^+ W^-$ production at the LHC}},
  \href{http://dx.doi.org/10.1103/PhysRevD.94.114014}{\emph{Phys. Rev. D} {\bf
  94} (2016) 114014}, [\href{http://arxiv.org/abs/1606.01034}{{\tt
  1606.01034}}].

\bibitem{Kallweit:2020gva}
S.~Kallweit, E.~Re, L.~Rottoli and M.~Wiesemann, \emph{{Accurate single- and
  double-differential resummation of colour-singlet processes with
  MATRIX+RADISH: $W^+W^-$ production at the LHC}},
  \href{http://dx.doi.org/10.1007/JHEP12(2020)147}{\emph{JHEP} {\bf 12} (2020)
  147}, [\href{http://arxiv.org/abs/2004.07720}{{\tt 2004.07720}}].

\bibitem{Bell:2020yzz}
G.~Bell, R.~Rahn and J.~Talbert, \emph{{Generic dijet soft functions at
  two-loop order: uncorrelated emissions}},
  \href{http://dx.doi.org/10.1007/JHEP09(2020)015}{\emph{JHEP} {\bf 09} (2020)
  015}, [\href{http://arxiv.org/abs/2004.08396}{{\tt 2004.08396}}].

\bibitem{BenekeFA}
M.~Beneke, \emph{Helmholtz International Summer School on Heavy Quark Physics,
  Dubna}, 2005.

\bibitem{Becher:2010tm}
T.~Becher and M.~Neubert, \emph{{Drell-Yan Production at Small $q_T$,
  Transverse Parton Distributions and the Collinear Anomaly}},
  \href{http://dx.doi.org/10.1140/epjc/s10052-011-1665-7}{\emph{Eur. Phys. J.
  C} {\bf 71} (2011) 1665}, [\href{http://arxiv.org/abs/1007.4005}{{\tt
  1007.4005}}].

\bibitem{Chiu:2011qc}
J.-y. Chiu, A.~Jain, D.~Neill and I.~Z. Rothstein, \emph{{The Rapidity
  Renormalization Group}},
  \href{http://dx.doi.org/10.1103/PhysRevLett.108.151601}{\emph{Phys. Rev.
  Lett.} {\bf 108} (2012) 151601}, [\href{http://arxiv.org/abs/1104.0881}{{\tt
  1104.0881}}].

\bibitem{Chiu:2012ir}
J.-Y. Chiu, A.~Jain, D.~Neill and I.~Z. Rothstein, \emph{{A Formalism for the
  Systematic Treatment of Rapidity Logarithms in Quantum Field Theory}},
  \href{http://dx.doi.org/10.1007/JHEP05(2012)084}{\emph{JHEP} {\bf 05} (2012)
  084}, [\href{http://arxiv.org/abs/1202.0814}{{\tt 1202.0814}}].

\bibitem{Becher:2012qa}
T.~Becher and M.~Neubert, \emph{{Factorization and NNLL Resummation for Higgs
  Production with a Jet Veto}},
  \href{http://dx.doi.org/10.1007/JHEP07(2012)108}{\emph{JHEP} {\bf 07} (2012)
  108}, [\href{http://arxiv.org/abs/1205.3806}{{\tt 1205.3806}}].

\bibitem{Becher:2009qa}
T.~Becher and M.~Neubert, \emph{{On the Structure of Infrared Singularities of
  Gauge-Theory Amplitudes}},
  \href{http://dx.doi.org/10.1088/1126-6708/2009/06/081}{\emph{JHEP} {\bf 06}
  (2009) 081}, [\href{http://arxiv.org/abs/0903.1126}{{\tt 0903.1126}}].

\bibitem{Gardi:2009qi}
E.~Gardi and L.~Magnea, \emph{{Factorization constraints for soft anomalous
  dimensions in QCD scattering amplitudes}},
  \href{http://dx.doi.org/10.1088/1126-6708/2009/03/079}{\emph{JHEP} {\bf 03}
  (2009) 079}, [\href{http://arxiv.org/abs/0901.1091}{{\tt 0901.1091}}].

\bibitem{Tackmann:2012bt}
F.~J. Tackmann, J.~R. Walsh and S.~Zuberi, \emph{{Resummation Properties of Jet
  Vetoes at the LHC}},
  \href{http://dx.doi.org/10.1103/PhysRevD.86.053011}{\emph{Phys. Rev. D} {\bf
  86} (2012) 053011}, [\href{http://arxiv.org/abs/1206.4312}{{\tt 1206.4312}}].

\bibitem{Cacciari:2008gp}
M.~Cacciari, G.~P. Salam and G.~Soyez, \emph{{The anti-$k_t$ jet clustering
  algorithm}},
  \href{http://dx.doi.org/10.1088/1126-6708/2008/04/063}{\emph{JHEP} {\bf 04}
  (2008) 063}, [\href{http://arxiv.org/abs/0802.1189}{{\tt 0802.1189}}].

\bibitem{Dokshitzer:1997in}
Y.~L. Dokshitzer, G.~D. Leder, S.~Moretti and B.~R. Webber, \emph{{Better jet
  clustering algorithms}},
  \href{http://dx.doi.org/10.1088/1126-6708/1997/08/001}{\emph{JHEP} {\bf 08}
  (1997) 001}, [\href{http://arxiv.org/abs/hep-ph/9707323}{{\tt
  hep-ph/9707323}}].

\bibitem{Wobisch:1998wt}
M.~Wobisch and T.~Wengler, \emph{{Hadronization corrections to jet
  cross-sections in deep inelastic scattering}},  in \emph{{Workshop on Monte
  Carlo Generators for HERA Physics (Plenary Starting Meeting)}}, pp.~270--279,
  4, 1998.
\newblock \href{http://arxiv.org/abs/hep-ph/9907280}{{\tt hep-ph/9907280}}.

\bibitem{Catani:1993hr}
S.~Catani, Y.~L. Dokshitzer, M.~H. Seymour and B.~R. Webber,
  \emph{{Longitudinally invariant $K_t$ clustering algorithms for hadron hadron
  collisions}},
  \href{http://dx.doi.org/10.1016/0550-3213(93)90166-M}{\emph{Nucl. Phys. B}
  {\bf 406} (1993) 187--224}.

\bibitem{Bauer:2020npd}
C.~W. Bauer, A.~V. Manohar and P.~F. Monni, \emph{{Disentangling observable
  dependence in SCET$_{I}$ and SCET$_{II}$ anomalous dimensions: angularities
  at two loops}}, \href{http://dx.doi.org/10.1007/JHEP07(2021)214}{\emph{JHEP}
  {\bf 07} (2021) 214}, [\href{http://arxiv.org/abs/2012.09213}{{\tt
  2012.09213}}].

\bibitem{Li:2016axz}
Y.~Li, D.~Neill and H.~X. Zhu, \emph{{An exponential regulator for rapidity
  divergences}},
  \href{http://dx.doi.org/10.1016/j.nuclphysb.2020.115193}{\emph{Nucl. Phys. B}
  {\bf 960} (2020) 115193}, [\href{http://arxiv.org/abs/1604.00392}{{\tt
  1604.00392}}].

\bibitem{Li:2016ctv}
Y.~Li and H.~X. Zhu, \emph{{Bootstrapping Rapidity Anomalous Dimensions for
  Transverse-Momentum Resummation}},
  \href{http://dx.doi.org/10.1103/PhysRevLett.118.022004}{\emph{Phys. Rev.
  Lett.} {\bf 118} (2017) 022004}, [\href{http://arxiv.org/abs/1604.01404}{{\tt
  1604.01404}}].

\bibitem{Huber:2005yg}
T.~Huber and D.~Maitre, \emph{{HypExp: A Mathematica package for expanding
  hypergeometric functions around integer-valued parameters}},
  \href{http://dx.doi.org/10.1016/j.cpc.2006.01.007}{\emph{Comput. Phys.
  Commun.} {\bf 175} (2006) 122--144},
  [\href{http://arxiv.org/abs/hep-ph/0507094}{{\tt hep-ph/0507094}}].

\bibitem{Duhr:2019tlz}
C.~Duhr and F.~Dulat, \emph{{PolyLogTools \textemdash{} polylogs for the
  masses}}, \href{http://dx.doi.org/10.1007/JHEP08(2019)135}{\emph{JHEP} {\bf
  08} (2019) 135}, [\href{http://arxiv.org/abs/1904.07279}{{\tt 1904.07279}}].

\bibitem{Li:2011zp}
Y.~Li, S.~Mantry and F.~Petriello, \emph{{An Exclusive Soft Function for
  Drell-Yan at Next-to-Next-to-Leading Order}},
  \href{http://dx.doi.org/10.1103/PhysRevD.84.094014}{\emph{Phys. Rev. D} {\bf
  84} (2011) 094014}, [\href{http://arxiv.org/abs/1105.5171}{{\tt 1105.5171}}].

\bibitem{Campbell:1997hg}
J.~M. Campbell and E.~W.~N. Glover, \emph{{Double unresolved approximations to
  multiparton scattering amplitudes}},
  \href{http://dx.doi.org/10.1016/S0550-3213(98)00295-8}{\emph{Nucl. Phys. B}
  {\bf 527} (1998) 264--288}, [\href{http://arxiv.org/abs/hep-ph/9710255}{{\tt
  hep-ph/9710255}}].

\bibitem{Dokshitzer:1997iz}
Y.~L. Dokshitzer, A.~Lucenti, G.~Marchesini and G.~P. Salam,
  \emph{{Universality of 1/Q corrections to jet-shape observables rescued}},
  \href{http://dx.doi.org/10.1016/S0550-3213(97)00650-0}{\emph{Nucl. Phys. B}
  {\bf 511} (1998) 396--418}, [\href{http://arxiv.org/abs/hep-ph/9707532}{{\tt
  hep-ph/9707532}}].

\bibitem{Catani:1999ss}
S.~Catani and M.~Grazzini, \emph{{Infrared factorization of tree level QCD
  amplitudes at the next-to-next-to-leading order and beyond}},
  \href{http://dx.doi.org/10.1016/S0550-3213(99)00778-6}{\emph{Nucl. Phys. B}
  {\bf 570} (2000) 287--325}, [\href{http://arxiv.org/abs/hep-ph/9908523}{{\tt
  hep-ph/9908523}}].

\bibitem{Kelley:2011ng}
R.~Kelley, M.~D. Schwartz, R.~M. Schabinger and H.~X. Zhu, \emph{{The two-loop
  hemisphere soft function}},
  \href{http://dx.doi.org/10.1103/PhysRevD.84.045022}{\emph{Phys. Rev. D} {\bf
  84} (2011) 045022}, [\href{http://arxiv.org/abs/1105.3676}{{\tt 1105.3676}}].

\bibitem{Monni:2011gb}
P.~F. Monni, T.~Gehrmann and G.~Luisoni, \emph{{Two-Loop Soft Corrections and
  Resummation of the Thrust Distribution in the Dijet Region}},
  \href{http://dx.doi.org/10.1007/JHEP08(2011)010}{\emph{JHEP} {\bf 08} (2011)
  010}, [\href{http://arxiv.org/abs/1105.4560}{{\tt 1105.4560}}].

\bibitem{Hornig:2011iu}
A.~Hornig, C.~Lee, I.~W. Stewart, J.~R. Walsh and S.~Zuberi, \emph{{Non-global
  Structure of the $\mathcal{O}({\alpha}^2_s)$ Dijet Soft Function}},
  \href{http://dx.doi.org/10.1007/JHEP08(2011)054}{\emph{JHEP} {\bf 08} (2011)
  054}, [\href{http://arxiv.org/abs/1105.4628}{{\tt 1105.4628}}].

\bibitem{Banfi:2014sua}
A.~Banfi, H.~McAslan, P.~F. Monni and G.~Zanderighi, \emph{{A general method
  for the resummation of event-shape distributions in $e^+e^-$ annihilation}},
  \href{http://dx.doi.org/10.1007/JHEP05(2015)102}{\emph{JHEP} {\bf 05} (2015)
  102}, [\href{http://arxiv.org/abs/1412.2126}{{\tt 1412.2126}}].

\bibitem{Beneke:1997zp}
M.~Beneke and V.~A. Smirnov, \emph{{Asymptotic expansion of Feynman integrals
  near threshold}},
  \href{http://dx.doi.org/10.1016/S0550-3213(98)00138-2}{\emph{Nucl. Phys. B}
  {\bf 522} (1998) 321--344}, [\href{http://arxiv.org/abs/hep-ph/9711391}{{\tt
  hep-ph/9711391}}].

\bibitem{Bell:2018vaa}
G.~Bell, R.~Rahn and J.~Talbert, \emph{{Two-loop anomalous dimensions of
  generic dijet soft functions}},
  \href{http://dx.doi.org/10.1016/j.nuclphysb.2018.09.026}{\emph{Nucl. Phys. B}
  {\bf 936} (2018) 520--541}, [\href{http://arxiv.org/abs/1805.12414}{{\tt
  1805.12414}}].

\end{thebibliography}\endgroup
\end{document}